\documentclass{article}

\usepackage[english]{babel}

\usepackage[letterpaper,top=2cm,bottom=2cm,left=2cm,right=2cm,marginparwidth=1.75cm]{geometry}

\usepackage{amsmath}
\usepackage{authblk}
\usepackage{graphicx}
\usepackage[numbers]{natbib}
\usepackage[colorlinks=true, allcolors=blue]{hyperref}
\usepackage{setspace}
\usepackage{caption}
\usepackage{float}
\usepackage{aliascnt}
\newaliascnt{eqfloat}{equation}
\newfloat{eqfloat}{h}{eqflts}
\floatname{eqfloat}{Equation}

\newcommand*{\ORGeqfloat}{}
\let\ORGeqfloat\eqfloat
\def\eqfloat{%
  \let\ORIGINALcaption\caption
  \def\caption{%
    \addtocounter{equation}{-1}%
    \ORIGINALcaption
  }%
  \ORGeqfloat
}

\author[*, a]{Rachel T Gonzalez}
\author[*, a]{Madeline R Abbott}
\author[b,c] {Brahmajee Nallamothu}
\author[b,c,d] {Scott Hummel}
\author[c,e]{Michael Dorsch}
\author[a]{Walter Dempsey}

\affil[a]{Department of Biostatistics, University of Michigan, 1415 Washington Heights, Ann Arbor, MI 48109, USA}
\affil[b]{Medical School, University of Michigan, 1301 Catherine Street, Ann Arbor, MI 48019, USA}
\affil[c]{Frankel Cardiovascular Center, University of Michigan, 1425 East Ann Street, Ann Arbor, MI 48019, USA}
\affil[d]{VA Ann Arbor Health System, 2215 Fuller Road, Ann Arbor, MI 48105, USA}
\affil[e]{College of Pharmacy, University of Michigan, 428 Church Street, Ann Arbor, MI 48109, USA}
\affil[*]{co-first authors}

\title{Practical considerations when designing an online learning algorithm for an app-based mHealth intervention}

\begin{document}
\maketitle

\noindent
corresponding author: Walter Dempsey \\
corresponding author email: wdem@umich.edu

\begin{abstract}
The ubiquitous nature of mobile health (mHealth) technology has expanded opportunities for the integration of reinforcement learning into traditional clinical trial designs, allowing researchers to learn individualized treatment policies during the study. \emph{LowSalt4Life 2} (LS4L2) is a recent trial aimed at reducing sodium intake among hypertensive individuals through an app-based intervention. A reinforcement learning algorithm, which was deployed in one of the trial arms, was designed to send reminder notifications to promote app engagement in contexts where the notification would be effective, i.e., when a participant is likely to open the app in the next 30-minute and not when prior data suggested reduced effectiveness.  Such an algorithm can improve app-based mHealth interventions by reducing participant burden and more effectively promoting behavior change.
We encountered various challenges during the implementation of the learning algorithm, which we present as a template to solving challenges in future trials that deploy reinforcement learning algorithms. We provide template solutions based on LS4L2 for solving the key challenges of (i) defining a relevant reward, (ii) determining a meaningful timescale for optimization, (iii) specifying a robust statistical model that allows for automation, (iv) balancing model flexibility with computational cost, and (v) addressing missing values in gradually collected data.  
\end{abstract}

\section{Introduction}

The ubiquitous nature of mobile health (mHealth) technology has enabled the integration of digital tools into clinical trials in a wide variety of domains, from substance use and mental health studies to cardiovascular disease and physical activity~\cite{10.1371/journal.pmed.1001362, klasnja_2019, rabbi_2018}.  These technologies not only enable high-frequency, low-burden data collection, but also help overcome obstacles in the delivery of care, making it possible to deliver behavioral interventions and provide accessible support to individuals anytime, anywhere.   
The primary goal of many digital intervention trials~\cite{klasnja_2015} is to assess whether and when it is useful to trigger delivery of behavioral interventions. These trials allow clinicians and behavioral scientists to build just-in-time adaptive interventions (JITAIs) that trigger behavioral interventions in the right moment and at the right time when individuals are most in need of support~\cite{nahum-shani_2018}. 

To determine when to deliver interventions, early digital intervention trials relied on treatment rules defined at baseline using relevant clinical knowledge~\cite{dorsch2018novel, dorsch2020effects}. In a recent smoking cessation study (Sense2Stop)~\cite{BATTALIO2021106534}, for example, the scientific team was most interested in understanding whether reminders to practice stress reduction exercises will be useful in reducing/preventing future stress if the reminders are delivered at times the participant is classified as stressed based on physiological measures from wearable sensors. The scientific team therefore decided to send some reminders at stress times and the remaining at times when the patient was not classified as stressed. A primary goal of this study was to assess whether the reminders, delivered at stress times, result in a reduction/prevention of stress over the subsequent hour and whether this effect changes with time.

While digital intervention trials such as Sense2Stop can lead to more effective intervention packages for future patients, more recent digital intervention trials~\cite{aguilera2020mhealth, liao2020personalized} have gone beyond rule-based protocols and focused on benefiting patients enrolled in the trial by learning personalized intervention policies, i.e., the chosen intervention at a given time may depend on the individual's past responses in contexts similar to their current context.  Building personalized intervention policies requires \emph{online reinforcement learning (RL) algorithms}~\cite{Sutton2018} that take as input streaming, real-time data and outputs a sequence of intervention rules that update throughout the trial.  These algorithms typically seek to optimize a measured outcome called a \emph{reward} and try to balance exploration -- the goal of exploring potential interventions that give high reward in a given context -- and exploitation -- the goal of choosing the optimal intervention that maximizes the reward in a given context.  

The most common online RL learning algorithms deployed in digital intervention trials are contextual bandit algorithms~\cite{Lattimore2020}.  These are an example of a myopic online RL algorithm, which takes as input the current context and outputs an intervention from the set of possible intervention options.  Canonical bandit algorithms include $\epsilon$-greedy, Upper Confidence Band, and Thompson Sampling~\cite{Lattimore2020}.   
Such algorithms were initially used in non-commercial, high-stakes domains such as adaptive clinical trials, where the algorithms were used to dynamically allocate patients to the most effective treatments to improve outcomes~\cite{jin2022seamless}. Over the past decade these algorithms have gained prominence in both marketing and e-commerce platforms, where they are used for real-time personalization, dynamic ad targeting, and adaptive pricing~\cite{Li2010, Bouneffouf2012}.
The application of contextual bandits to mHealth research is much more recent.  Thompson Sampling is the most common approach in digital intervention trials due to its stochastic nature, whereas Epsilon-Greedy or Upper Confidence Bound (UCB) use deterministic rules to choose actions.  Thompson Sampling maintains a probability distribution for the expected reward of each intervention option and, at each step, samples from these distributions to select the intervention with the highest sampled value.  The probabilistic algorithm underlying Thompson Sampling (with minor modifications) ensures clinicians and behavioral scientists can perform statistical inference post-study~\cite{NEURIPS2024_e7b3e34d,zhang2025replicablebanditsdigitalhealth}, allowing the study team to assess whether and when it was useful to trigger delivery of behavioral interventions. 

In this paper, we present a digital intervention trial that uses a contextual bandit algorithm to optimize app-based notifications aimed at addressing high dietary sodium intake among individuals with hypertension.  When designing the statistical algorithm embedded in the digital intervention trial, the study team had to balance statistical complexity and computational cost --- a task that required making many decisions motivated by practical concerns.  We present these challenges and our solutions as a template to solving challenges in future trials that deploy online RL algorithms. 

\subsection{Related work}
In tandem with the growing application of contextual bandits in mHealth, researchers have considered best practices for designing and implementing trials that incorporate such components into their protocols. We mention a few works providing case-studies and guidance here.

Figueroa et al. \cite{figueroa2021adaptive} addressed challenges that emerged throughout the execution of the DIAMANTE study, an randomized controlled trial (RCT) that aims to increase physical activity in participants through text messaging interventions. One arm of the three-arm trial employs RL techniques to personalize message delivery. They summarize challenges into categories relating to model building, data handling and collection, and balancing algorithm performance with practical implementation constraints. In addition to their careful review of emergent issues, the authors also discuss their solutions in the context of DIAMANTE. Our current work builds on their list of challenges by focusing on statistical concerns in designing online reinforcement learning algorithms. Specifically, our paper will discuss statistical issues of model complexity and robustness and how we solved them with data-adaptive model selection and weakly informative priors.  

A more general discussion of frameworks for RL algorithm development in mHealth is provided by Trella et al.\cite{Trella_2022}. They extend a general framework for the assessment of prediction algorithms to the design of mHealth MRTs with reinforcement learning components. Their approach has three pillars: personalization, the degree to which an algorithm can learn to take the best action for each user; computability, the ability of an algorithm to run efficiently online; and stability, the capacity of an algorithm to operate mostly autonomously and in a wide variety of situations. This framework is illustrated in depth through its application to a digital intervention trial in oral health, Oralytics.  Our current work builds on their general framework by diving into the technical tradeoffs of these three pillars. In particular, our data-adaptive approach to model building nicely handles the personalization/computation/stability trade-off.  Moreover, our use of weakly informative default priors~\cite{Gelman2008} ensures our algorithms will always give answers, even when there is complete separation in logistic regression.  Separation is a common problem even when the sample size is large and the number of predictors is small, and can plague na{\"i}ve choice of default Gaussian or Laplace priors. We will show how this approach automatically applies more shrinkage to higher-order interactions, allowing for personalized intervention rules while ensuring robust algorithmic performance.

\subsection{Outline}

Researchers attempting to implement a contextual bandit in a digital intervention trial will encounter various challenges, which we categorize into five general areas: (i) how to define a relevant reward, (ii) how to determine the timescale of optimization, (iii) how to specify a robust statistical model that allows for automation, (iv) how to balance model flexibility with computational cost, and (v) how to address missing values present in data that are collected gradually.  Here, we describe important factors to consider when developing solutions to these challenges and outline pragmatic solutions to these general practical problems using \emph{LowSalt4Life 2} as a running example.  This paper aims to serve as a template for thinking about these potential challenges, which we believe will be encountered in similar studies involving contextual bandit algorithms in mHealth.  We present the motivating trial in Section \ref{s:mrt_design}, describe commonly encountered practical issues in Section \ref{s:practical_issues}, illustrate our pragmatic solution in Section \ref{s:simulations} through simulation, and conclude with a discussion in Section \ref{s:discussion}.

\section{Motivating trial design: \emph{LowSalt4Life 2}}\label{s:mrt_design}

\begin{figure}
    \centering
    \includegraphics[width=0.7\linewidth]{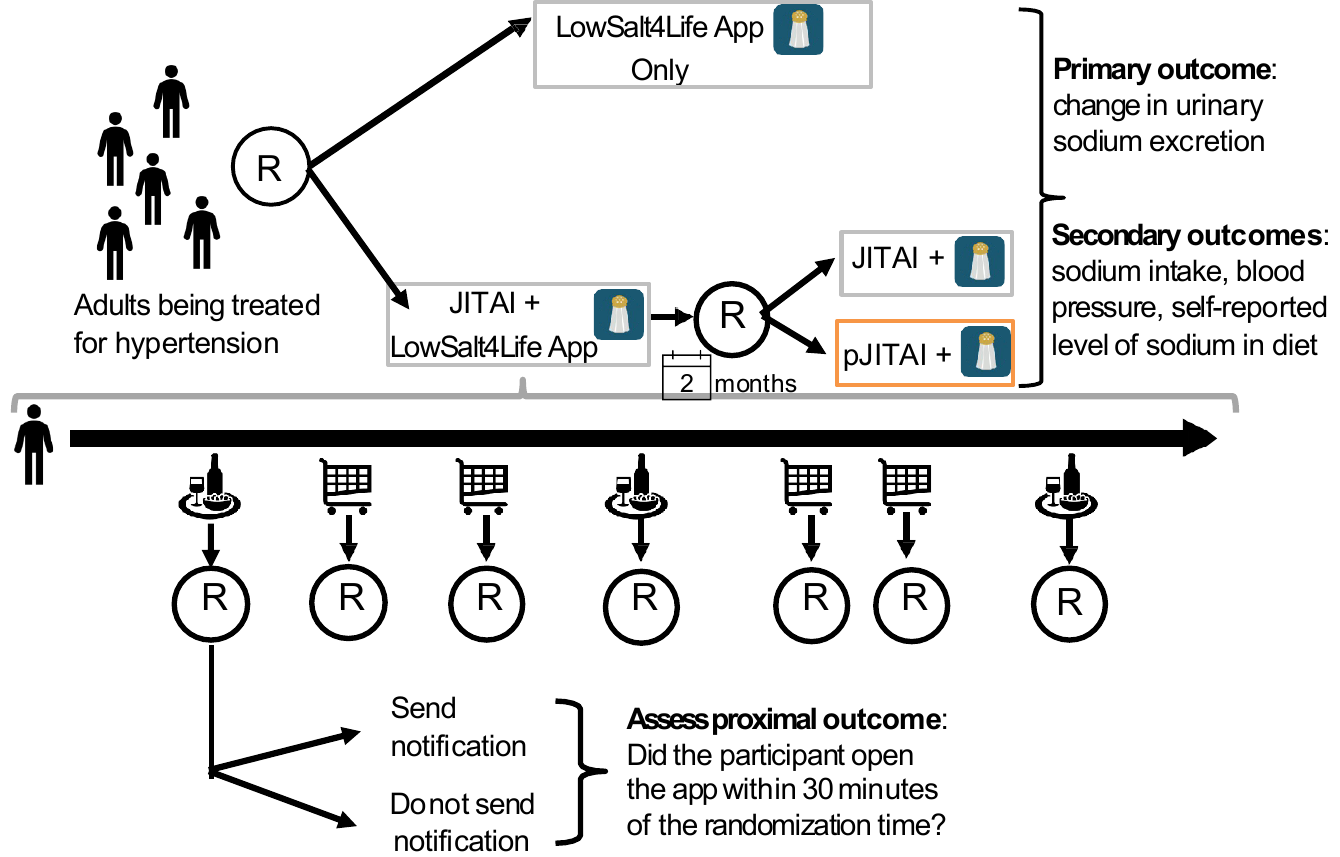}
    \caption{LS4L2 is a two-arm (App Alone vs App + JITAI) trial. After two months, participants in the App + JITAI arm undergo a second randomization to either continue with the current JITAI or transition to a personalized JITAI (pJITAI) . The App + pJITAI sub-arm includes a contextual bandit algorithm to optimize delivery of an app-based notification.  Within this sub-arm, participants are randomized to potentially receive a mobile notification from the study's app each time they arrive at a grocery store or restaurant.  The randomization probabilities are based on context covariates, all of which are collected passively.}
    \label{fig:trial_design}
\end{figure}

The primary objective of the \emph{LowSalt4Life 2} (LS4L2) trial is to evaluate whether the inclusion of a JITAI within a mobile application (App + JITAI) reduces dietary sodium intake more than the mobile application alone (App Alone) among hypertension patients.  This trial is a two-arm RCT in which adults currently being treated for hypertension are randomized with equal probability to receive or not receive a JITAI through a mobile app.  The JITAI is triggered based on arrival at a grocery store or restaurant and takes as input the current context of the participant.  During the first 2 months in the study, participants are then randomized to receive an app-notification with probability 80\%, and not receive an app-notification with probability 20\%.  The notifications aim to increase proximal app engagement.
At 2 months, participants in the JITAI arm are randomized again with equal probability to either continue with the current JITAI or transition to a personalized JITAI (pJITAI). A diagram of the trial is given in Figure \ref{fig:trial_design}. Details of the LS4L2 protocol have been published previously \cite{dorsch2025}.
For the purposes of this paper, we focus on the development of the contextual bandit algorithm used only in the App + pJITAI arm of the trial in which participants receive the pJITAI.  

Since the US population tends to consume most of its dietary sodium from processed foods and foods eaten at restaurants \cite{cdc_2023}, the JITAI was designed to target participants when they arrive at a grocery store or restaurant by using these events as decision points within the App + JITAI trial arm.
Arrival at either location is passively-sensed using geo-fencing and GPS. We refer to these times as \emph{decision times}. At decision times, the mobile phone provides the following passively-collected context variables:  the time of the week (e.g., weekend, weekday, holiday), time of the day (e.g., morning, afternoon, etc.), current situation as coded by the mobile system  (e.g., shopping, social, working out, etc.), weather (e.g., very cold, cold, warm, etc.), and past app engagement (1 if the participant has opened the app in the past week and 0 otherwise).  In LS4L2, all measured context variables are categorical.
Within the App+pJITAI sub-arm, the Thompson sampling algorithm consists of two components.  First, at every decision time~$t$ and for every participant~$i$ in the study, there is a known \emph{probabilistic mapping}~$\pi_{i,t}: \mathcal{X} \to [0,1]$ that takes as input the current context~$X_{i,t} \in \mathcal{X}$ for that participant (i.e., set of current categorical variables) and outputs a \emph{probability} of sending an app-based notification or not (denoted $\pi_{i,t}(X_{i,t})$).  Second, there is a \emph{learning algorithm} that takes the current collected data as input and learns the probabilistic mapping. The goal of the Thompson sampling algorithm is to optimize delivery of the notifications to be at times when the participant is more likely than not to engage with the study app.   In LS4L2, we use a type of logistic regression to learn the mapping from a given context to the context-specific probability of sending a notification. We will discuss each subcomponent of the algorithm in detail in the sections below.

\section{Commonly encountered practical issues}\label{s:practical_issues}

In this section, we present practical issues encountered when designing the contextual bandit algorithm used to optimize the JITAI within the App+pJITAI arm in the LS4L2 trial.

\subsection{Constructing a relevant reward for the learning algorithm}

RL algorithms, such as contextual bandit algorithms, require a pre-specified reward. These algorithms then use this reward as a measurable signal to decide in which contexts it is appropriate to provide an intervention or stimulus (e.g., app-based notification) and in which contexts it is not.
From the perspective of designing interventions for behavior change, the scientific goal is to use a stimulus (e.g., notification from the app) to encourage participants to engage in a certain behavior (e.g., thinking about dietary sodium consumption and choosing low sodium alternatives).  In mHealth, measuring the desired behavior moments after the Thompson Sampling algorithm makes a decision is difficult.  While tools for real-time measurements such as ecological momentary assessments (EMAs)~\cite{shiffman2008ecological} have been used to collect in-the-moment data, the ability to passively detect the desired behaviors is often either impossible or not yet available due to current limitations in sensing technologies.  In LS4L2, the scientific team decided that EMAs had three disadvantages: (1) self-report was too high burden, (2) EMA could result in a high degree of missingness, and (3) self-reported behavior may not be a good measure of underlying behavior. 
Lab-based measures of dietary sodium intake were deemed not feasible due to the need for proximal measures of the desired behavior.  Finally, there are no passively collected proximal measures of the desired behavior.  

The inability to measure the task of interest is a common problem in digital intervention trials.  A common issue in the development of online RL algorithms for behavior change is the conflation of the algorithm's reward and the behavioral task of interest.  We solve this practical challenge by operationalizing a relevant, passively measurable reward that serves as a suitable proxy for the desired behavior.  By viewing the reward as a proxy, the study team can assess its suitability for learning good policies aimed at getting participants to engage in the behavior of interest.  In LS4L2, the scientific team decided that a binary outcome based on whether or not the user engages with the app in the 30-minute window following the decision time is a suitable proxy for our proximal behavioral outcome of true interest---whether or not someone makes low sodium choices.  Interacting with the app may include searching for low-salt foods at a given restaurant or looking up alternative low-salt options when purchasing food at a grocery store.  The choice of the 30-minute window attempts to strike a balance between signal and noise.  That is, we aim to allow enough time to detect the participant's possible interaction with the app after arriving at a grocery store or restaurant (i.e., after the decision point) while minimizing potential noise included in our measurement of the reward if we were to wait an extended period of time to observe a potential app interaction.  

It is critical for scientific teams designing bandit algorithms to think about whether the reward is a suitable proxy for the behavior of interest. In the first few months after study enrollment, thinking of and choosing low sodium alternatives was not thought to be part of the participant's routine when entering grocery stores or restaurants.  Therefore, proximal app engagement was seen to be a good proxy for whether the participant practiced the behavior of interest.  In the final month of the study, if the intervention was successful at helping a participant achieve behavior change, that participant may be less likely to use the app and may successfully practice the behavior without app support.  In this case, the notification may still serve as a reminder and be helpful in maintaining the desired behavior; however, proximal app engagement will no longer be a good proxy. For this reason, this choice of reward may lead the RL algorithm to send fewer messages than desired in the later stages of the study.  Due to habituation and burden, the study team found this a suitable tradeoff.   After LS4L2 is completed, a critical analysis will try to better understand whether the proxy reward was suitable and how to improve our choice for future studies.  Our solution to the challenge of identifying a relevant reward by disentangling the reward from the stimulus/behavior framework makes such analyses possible.

\subsection{Determining a meaningful timescale for optimization}

Recall that the RL algorithm consists of two subroutines: (1) a probabilistic mapping~$\pi_{i,t}$ for every participant at a given decision time and (2) a learning algorithm that builds the probabilistic maps using prior data.  Because participants are enrolled in the study sequentially, the amount of potential information available to the learning algorithm gradually increases as the study progresses.  As a result, little information will be available to the learning algorithm for the first few participants enrolled in the study, but a larger information base will be available for participants enrolled later in the study.   In the design of a contextual bandit algorithm, there are three different timescales that are relevant: the reward timescale (a 30-minute window after randomization), the decision timescale of the decision points (frequency of trips to grocery stores or restaurants), and the learning timescale (updates to the probabilistic maps used to deliver notifications). A key consideration becomes the appropriate timescale for the learning algorithm. In other words, how often is it appropriate to update our mapping~$\pi$ from the contextual environment to the randomization probability for each study participant? Our process for making these determinations is described below.

In LS4L2, we identified two considerations that guided this choice.  First, the reward and decision point timescales can inform the appropriate learning timescale.  Based on historical data, we expected only a few (2-4) decision points every week per individual.  Since the reward was binary, we wanted to ensure sufficient data so that the learning algorithm would construct probabilistic maps that were different from the previous ones.  Second, we needed to consider the expected monetary cost of computation per participant in the study.  Fitting the hierarchical Bayesian logistic regression model that operates behind the scenes in our learning algorithm can require substantial computing power, so we must consider what the total cost will be if we must fit the model multiple times to each participant.  Although we might expect a significant behavior change at the month-by-month level, doing so would be too expensive given the total number of participants in the study.   Conducting the first update after two months facilitates exploration and accumulation of data about a binary behavior (i.e., opening the study app) observed at relatively infrequent decision points (i.e., a few times per week).  The scientific team decided that two months is a sufficient time period in which we expect to observe any potential meaningful behavior changes, such as habituation to app notifications. Therefore, in order to balance the gradual accumulation of potential information useful for optimization with the computational cost of carrying out an update to the delivery rules, the learning timescale was set to every 2 months, so that the learning algorithm was run at 2 and 4 months after recruitment for every participant to update their probabilistic maps.

\subsection{Building a robust, automated learning algorithm}

Contextual bandit algorithms enable personalized probabilistic maps to be built over time during the course of the study. In clinical trials, it is imperative that the learning algorithm is feasibly deployed in practice, so that repeated updates of the learning algorithm (in months 2 and 4 since recruitment for every participant) neither ``break'' (i.e., fail to construct a new probabilistic map) nor require monitoring/maintenance that was not part of the study protocol. In LS4L2, the learning algorithm consists of fitting a hierarchical Bayesian logistic regression model with both population-level and individual-level parameters and reporting summaries based on the resulting posterior distributions.  In LS4L2, we define a model as ``breaking'' if the Bayesian regression either fails to run (i.e., any error resulting in no posterior draws) or does not produce reasonable posterior distributions (i.e., issues with mixing of MCMC chains).   Such outcomes are possible for Bayesian logistic regression when either the model (a) is fit to a small volume of data and/or (b) has too many predictors or interaction terms included.  Here, we discuss our approach to ensure robust learning and establish a protocol for automated monitoring of the learning algorithm.

To make our algorithm robust when we have both small amounts of data (i.e., at the start of the study, since data are collected sequentially) and higher order interaction terms (which may be important predictors of response to a notification), we use a combination of partial pooling and shrinkage priors. Specifically, our Bayesian logistic regression model is designed to be hierarchical (i.e. includes patient-specific parameters), enabling partial pooling of information.  Partial pooling allows the probabilistic map for a participant to start at a population-level mapping when there are few participant-level data and become more personalized through the inclusion of participant-specific parameters as data accumulate.  Even with partial pooling, a common issue for Bayesian logistic regression is \emph{separation}, which occurs when a context variable perfectly predicts the binary reward.  To ensure model stability and avoid issues of separation, we employed weakly informative default priors~\cite{Gelman2008} that are designed to have higher degrees of shrinkage as the complexity of the term increases.  Specifically, the prior on higher order interaction terms are designed to have a significantly larger amount of shrinkage towards zero than main effects.  In LS4L2, the same shrinkage priors are used for both population-level and individual-level model parameters. 

\subsubsection{Algorithm Monitoring: A new component of study monitoring for digital intervention trials}

Study monitoring is a critical component of clinical trials, as it ensures the integrity, quality, and accuracy of the data collected. Protocols for monitoring data collection are essential because they provide a detailed roadmap for how data will be collected, managed, and monitored throughout the study. Adhering to a robust monitoring plan helps to identify and resolve issues early, prevent protocol deviations, and ultimately ensure the reliability of the study's findings.  When deploying bandit algorithms, it is important to also draft a study protocol for monitoring of the algorithm in addition to standard trial protocols. These algorithms require two types of monitoring systems: (1) a \emph{failure monitoring system} to monitor failure of the learning algorithm to update and failure of the probabilistic mappings to work properly, and (2) a \emph{statistical monitoring system} to monitor whether the probabilistic mappings are working as intended.

In LS4L2, the failure monitoring system consists of checks on the output of the learning algorithm and whether the per-participant updates were running at the specified times (2 and 4 months since recruitment).  The study team downloads the output of the learning algorithm from the cloud and runs sanity checks by comparing to locally fit versions of the model.  This allows the study team to identify potential software bugs or issues with the collected data. Any bug/issue is (1) dated, (2) documented as a protocol deviation, and (3) the software changes are recorded in a log file.  Our failure monitoring was able to identify a key issue in data collection due to software updates that was remedied within days.  The issue was fixed and documented as a protocol deviation.  Without such failure monitoring systems, mHealth clinical trials will only identify these failure modes after the trial is completed.

\begin{figure}
    \centering
    \includegraphics[width=0.7\linewidth]{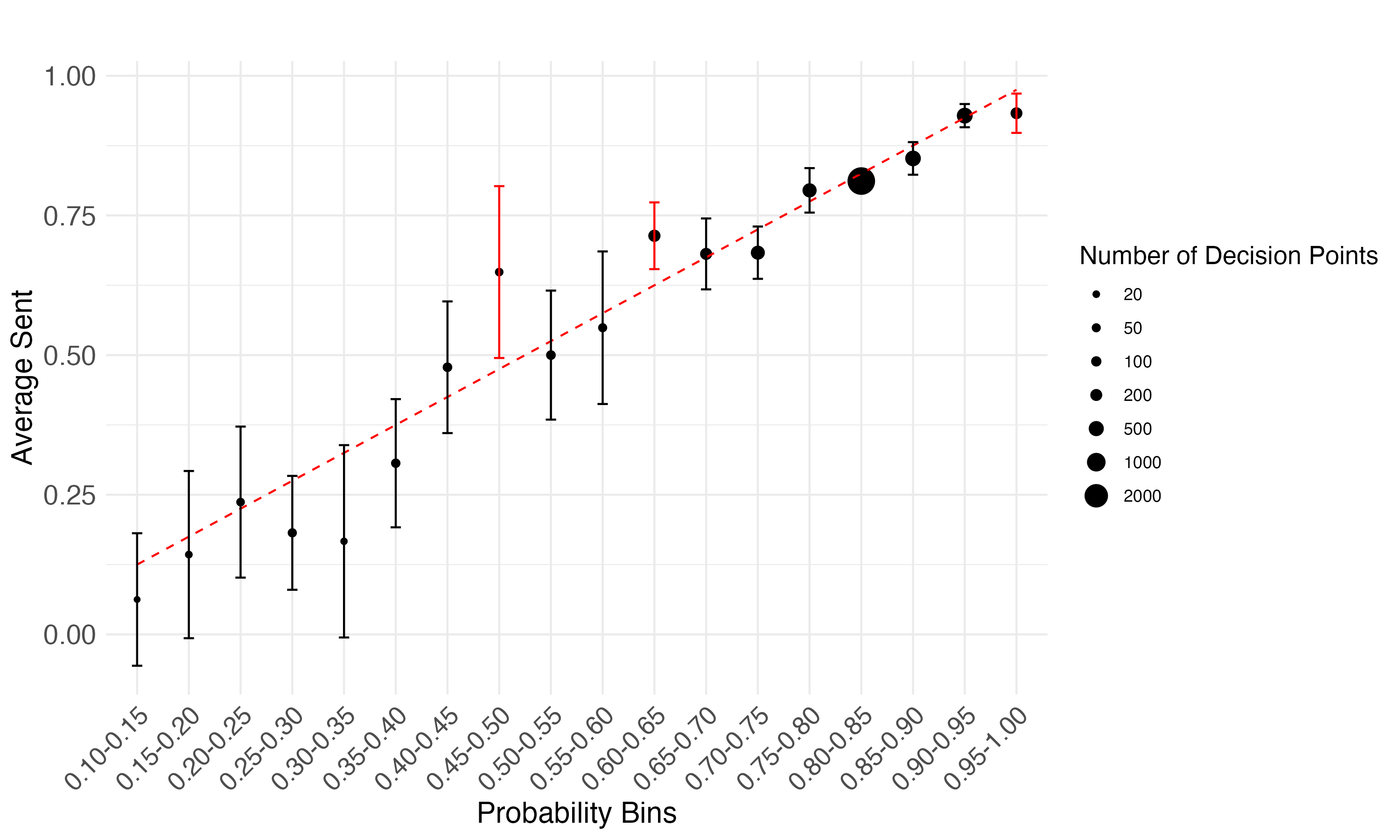}
    \caption{As an additional layer of study monitoring, we examined the calibration of our algorithm. We calculated the empirical probability with which notifications were sent to study participants in bins corresponding to the individualized decision rules learned by the learning algorithm. Uncertainty was quantified with asymptotically normal 95\% CIs for the observed probabilities.}
    \label{fig:calibration_plot}
\end{figure}

In addition, the study team decided on two statistical checks to ensure fidelity to the algorithm protocol.  First, we assessed whether the MCMC chains stored on the cloud matched closely with local runs of the same updates.  Automated reports were generated for every participant at each time that the learning algorithm was run.  A data analyst reviewed these documents for potential deviations/bugs.  In studies with more frequent updates, such reports may not be able to be manually reviewed and an automated review should be protocolized.  Second, we built a statistical monitoring system for the probabilistic maps.  Specifically, we assessed the statistical calibration of our algorithm. Figure \ref{fig:calibration_plot} summarizes our approach. We categorized all decision points by the personalized and context-specific randomization probabilities. In each bin of width 0.05, we then calculated the empirical probability with which notifications were actually sent and a 95\% asymptotically normal confidence interval. If our implementation is well calibrated then we would expect the confidence intervals to cover the midpoint of each bin. In Figure \ref{fig:calibration_plot}, LS4L2 is well calibrated with three bins with statistically significant deviation which the study team deemed acceptable.

\subsection{Balancing model flexibility and computational cost}  

In the contextual bandit literature it is common to assume correct model specification.  In practice, the model form is not known, which leads to the natural question of how to balance the complexity of the model with the amount of information currently available.  Consider LS4L2 and its set of categorical context variables. One potential model (Model A) would include all categorical variables and an intervention indicator.  Another potential model (Model B) would include all categorical variables, their pairwise interactions, the interaction of each categorical variable with an intervention indicator, and the three-way interaction of two categorical variables and intervention. If we have 5 binary context variables and 1 intervention indicator, then Model A would have 6 parameters (5 main effects, 1 intervention effect) while Model B would have 50 parameters (5 main effects, 20 pairwise interactions, 5 main intervention effects, and 20 pairwise intervention interactions).  Model A may be too simple and ignore contextual moderators; however, Model B may be too complex and lead to slower learning of important contextual moderators.  Model B may also suffer from instability early in the study when little data has been collected. Moreover, there is substantial difference in computational cost between Models A and B.  

\begin{figure}
    \centering
    \includegraphics[width=0.8\textwidth]{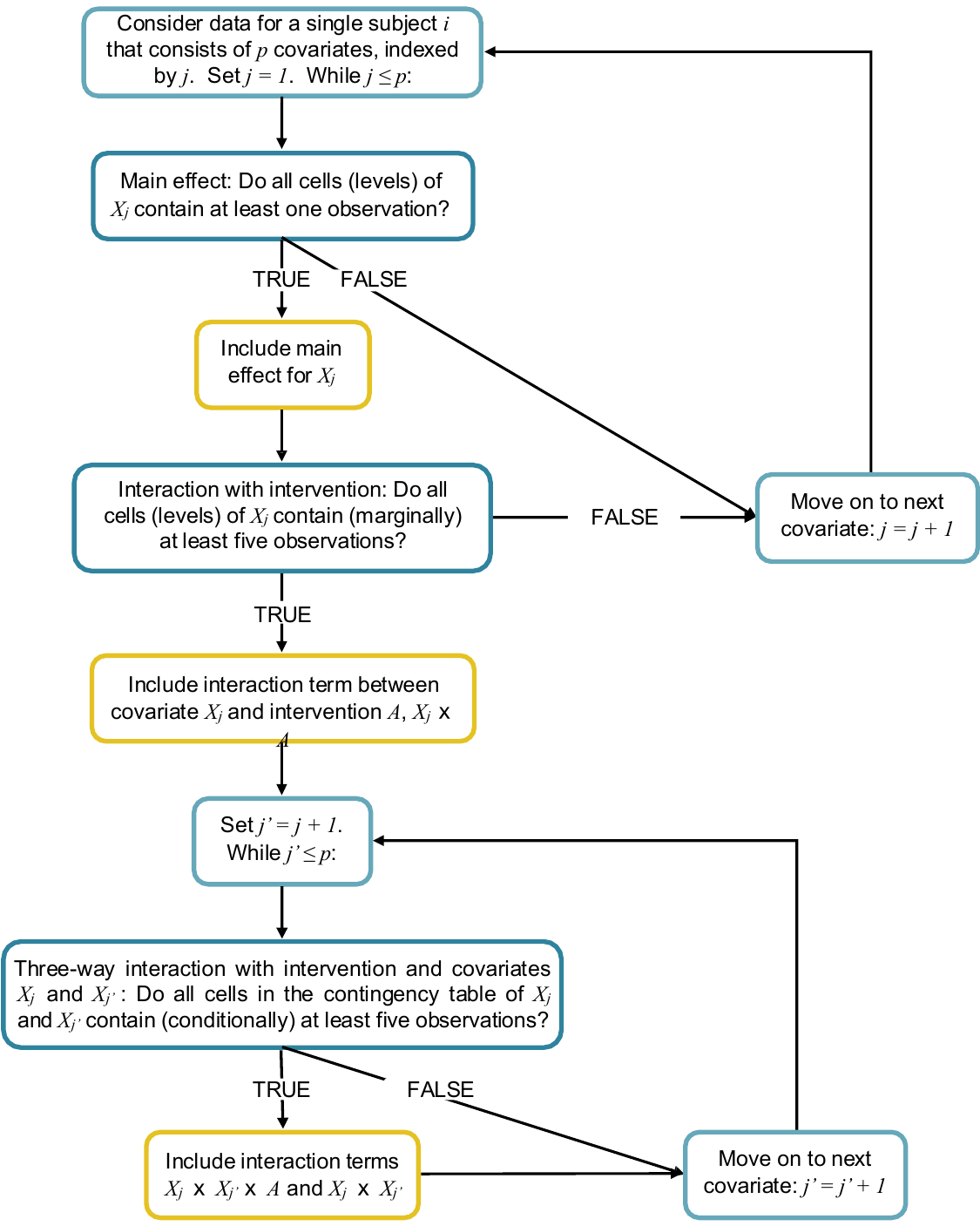}
    \caption{Determining model complexity for the single participant model.  Assume that the data for the current participant contains $p$ covariates (excluding intervention), which we index by $j$.  We apply these rules to determine inclusion of fixed effects.  Note that the number of ``unknown" observations for each covariate are not counted towards the minimum cell size.  For inclusion of random effect, a similar set of rules is used by only main effects and first-order interactions between intervention and covariates are considered (no three-way interactions).  Different rules are applied for inclusion of baseline covariates.}
    \label{fig:mod_flow_chart}
\end{figure}

To directly address this trade-off, the LS4L2 study team decided that model specification would take into account observed variation in the categorical variables while respecting an upper limit on the number of interaction terms in the model. Figure~\ref{fig:mod_flow_chart} visualizes this process for the context variables, showing how the study team balanced model complexity and computational cost with the amount of information available in the data. 

In addition to intervention personalization with respect to context variables, LS4L2 accounts for potential differences in responsiveness to notifications over the course of the study (e.g., habituation) by incorporating an indicator variable for each study period.  These study period binary indicators are denoted $S_1$ and $S_2$, and ``turn on" after the 2- and 4-month updates respectively (i.e, $S_1 = \textbf{I}(\texttt{daysInStudy} > 56)$ and $S_2 = \textbf{I}(\texttt{daysInStudy} > 112)$).  In addition to Figure~\ref{fig:mod_flow_chart}, the LS4L2 team defined rules for including study period-specific interaction terms using the rules as described above but within each study period.  That is, the first round applies the rules for including model terms to all the data, the second round applies the rules for including model terms to only data collected after the 2 month optimization point (i.e., when $S_1 = 1)$), and (3) the third round applies the rules for including model terms to only data collected after the 4 month optimization point (i.e., when $S_2 = 1)$).  All terms added to the model in the second and third rounds are interacted with $S_1$ and $S_2$, respectively. For baseline categorical covariates (i.,e., age and gender), we apply a slightly different set of rules to determine their inclusion.  These rules---for including both main effects and interaction terms---are based on the variability in patient age and gender (See Appendix section \ref{sec:rules} for more details).

\subsection{Addressing missing values in gradually collected data}

Missing data are commonplace in clinical trials but presents an additional challenge when designing online RL algorithms \cite{little2012prevention}.
Distinct challenges arise in both subroutines of the RL algorithm (i.e., probabilistic mapping and learning algorithm).  

Recall that the learning algorithm requires fitting a Bayesian hierarchical logistic regression model.  In LS4L2, the reward and intervention are always observed, but certain categorical contextual variables may be missing.  For contextual variables with missing data, the LS4L2 team decided to augment the existing categories with a new level termed ``Unknown".  Model fitting via MCMC is then performed with this strategy.  In other digital intervention trials, the context will likely include non-categorical variables. One approach is to consider predicted context and build a measurement error model as in Guo, Xu, and Murphy \cite{pmlr-v238-guo24b}.  However, it is often the case that the measurement error associated with predicted contexts is not well understood.  Our practical recommendation is to build context variables via a well documented protocol, and have criteria for when a variable is deemed ``missing''. Under these criteria, a strategy similar to that used in LS4L2 is feasible even for non-categorical context variables.  Specifically, we include a binary indicator of missingness as part of the context and use it when the context variable is missing, and use the observed value when the context is not missing.

For the probabilistic mappings, a key challenge that arises with categorical context variables is that there will be contexts that have never been observed for which we may need to generate a probabilistic mapping.  Consider, for example, a study that starts recruitment in the summer and uses a categorical context variable to summarize the current weather.  One potential level of this variable is ``Cold''.  However, in early stages of the study, there will be many updates of the Bayesian hierarchical logistic regression which will not have access to data that includes the level ``Cold''.  In this case, the in-the-moment decision making algorithm may request a probability of sending a notification in a context that includes ``Cold'' but the probabilistic map cannot be built using the regression model fit by the learning algorithm. In LS4L2, we brake up these types of scenarios into two categories:

\begin{itemize}
    \item {\bf Scenario 1} is when the context variable is measured as ``unknown'', but the level ``unknown'' for that variable is not included in the data used to fit the model.  We refer to Scenario 1 as the \emph{missing data scenario}.
    \item {\bf Scenario 2} is when the context variable is measured as any level except ``unknown'', but the measured level is not included in the data used to fit Bayesian hierarchical logistic regression. We refer to Scenario 2 as the \emph{new context prediction scenario}.
\end{itemize}

Both settings require rules for propagating the updated intervention probabilities from the fitted Bayesian hierarchical logistic regression to the scenarios described here. In LS4L2, the study team considers only categorical context variables.  This meant that all potential contexts can be enumerated in a large table (each row representing a different context combination) and those combinations falling under Scenario 1 and 2 can be flagged.  For rows that are not flagged, the probabilistic map can be constructed using the Bayesian hierarchical logistic regression and a column is added to the table to represent the probability of sending a notification in this context.  For rows that are flagged, we implement a procedure for building the probability map that depended on the scenario.  In the missing data scenario (Scenario 1), we impute a value for the updated intervention probability by taking an average across all available levels of the covariate in the small table (the subset of the large table with only unflagged rows). In the new context prediction scenario, we impute a value for the updated intervention probability by either averaging across all available levels of the context variable or linking the unavailable level of the context variable to the closest value for which an intervention probability is available. For example, if a probability is unavailable for ``cold'' temperatures but it is available for ``cool'' temperatures, then temporarily set all levels of ``cold'' to ``cool''.

\section{Simulation Study}\label{s:simulations}

\subsection{Simulation Design}

To illustrate the advantage of our approach, we study our LS4L2 algorithm and its performance relative to two alternative contextual bandit learning algorithms that were considered when designing the LS4L2 clinical trial. Both alternative algorithms utilize Bayesian logistic regression models to update the probabilistic mapping but differ in their degree of model complexity. As defined previously, let $\boldsymbol{X} = (X_1, ..., X_p)$ denote the set of context variables and let $A$ be a binary variable that indicates whether or not a participant received the app-based intervention. The first algorithm, referred to as ``Simple'', relies on the model shown in Equation \eqref{e:simple} with main effects for the intervention indicator and context covariates only. The second, referred to as ``Complicated'', estimates the probability that the app is opened using the maximal model in the LS4L2 learning algorithm (see Equation \eqref{e:complicated}).

\begin{eqfloat}
\begin{equation}
\begin{aligned}
    \text{logit}(p_i|A_i, X_i) &= \beta_0 + \beta_1 A_i + \sum_{j = 1}^{p} \beta_{2j} X_{ij} \\
    &\quad + \left[\beta^{(1)}_0 + \beta^{(1)}_1 A_i + \sum_{j = 1}^{p} \beta^{(1)}_{2j} X_{ij}\right] \times S_{1i} \\
    &\quad + \left[\beta^{(2)}_0 + \beta^{(2)}_1 A_i + \sum_{j = 1}^{p} \beta^{(2)}_{2j} X_{ij}\right] \times S_{2i}
\end{aligned}
\label{e:simple}
\end{equation}
\caption{Logistic regression model for learning the probability of opening the study app at each decision point when the study participant did/did not receive the intervention in a specific context in the ``Simple'' learning algorithm.}
\end{eqfloat}

\begin{eqfloat}
\begin{equation}
\begin{aligned}
    logit(p_i|A_i, X_i) = \beta_0 + \beta_1 A_i + \sum_{j = 1}^{p} \beta_{2j} X_{ij} + \sum_{j = 1}^{p} \sum_{j^{'} = j+1}^{p} \beta_{3j,j^{'}} X_{ij} X_{ij^{'}} + \sum_{j = 1}^{p} \beta_{4j} A_i X_{ij} + \sum_{j = 1}^{p} \sum_{j^{'} = j+1}^{p} \beta_{5j,j^{'}} A_i X_{ij} X_{ij^{'}} \\
    + \bigg[\beta^{(1)}_0 + \beta^{(1)}_1 A_i + \sum_{j = 1}^{p} \beta^{(1)}_{2j} X_{ij} + \sum_{j = 1}^{p} \sum_{j^{'} = j+1}^{p} \beta^{(1)}_{3j,j^{'}} X_{ij} X_{ij^{'}}  +\sum_{j = 1}^{p} \beta^{(1)}_{4j} A_i X_{ij} + \sum_{j = 1}^{p} \sum_{j^{'} = j+1}^{p} \beta^{(1)}_{5j,j^{'}} A_i X_{ij} X_{ij^{'}} \bigg] \times S_{1i} \\
    + \bigg[\beta^{(2)}_0 + \beta^{(2)}_1 A_i + \sum_{j = 1}^{p} \beta^{(2)}_{2j} X_{ij} +  \sum_{j = 1}^{p} \sum_{j^{'} = j+1}^{p} \beta^{(2)}_{3j,j^{'}} X_{ij} X_{ij^{'}} + \sum_{j = 1}^{p} \beta^{(2)}_{3j} A_i X_{ij} + \sum_{j = 1}^{p} \sum_{j^{'} = j+1}^{p} \beta^{(2)}_{5j,j^{'}} A_i X_{ij} X_{ij^{'}} \bigg] \times S_{2i}
\end{aligned}
\label{e:complicated}
\end{equation}
\caption{Logistic regression model for learning the probability of opening the study app at each decision point when the study participant did/did not receive the intervention in a specific context in the ``Complicated'' learning algorithm.}
\end{eqfloat}

We consider two settings for the true distribution of the binary reward given the context covariates. In setting 1, we assume that the true effect of the intervention is not moderated by context covariates. The model governing the true probability of opening the app in a given context is given in Appendix Equation \eqref{e:s1}. Details on how we chose the model coefficients can be found in Appendix Table \ref{t:interactions} with exact coefficient values listed in \texttt{setting1\_coefficients.csv}. We designed this setting to align with the ``Simple'' algorithm, but include several additional two-way interactions between context covariates so that the logistic regression underlying the ``Simple'' algorithm is misspecified.

\noindent In setting 2, we assume the reward distribution is characterized by a high degree of intervention effect heterogeneity. The generative model is similar to Appendix Equation~\eqref{e:s1} but includes additional interactions between the intervention indicator and each of the contextual moderators in LS4L2.  In fact, the generative model in this setting exactly matches the assumptions of the ``Complicated'' algorithm. We design the simulation so that time of the week, time of day, and app engagement are strong moderators.  These choices align with the expectations of our study team regarding effect moderators in mHealth clinical trials. In addition, the effect of treatment is designed to decrease over the course of the study, but remain positive. The detailed specification of the generative model for setting 2 is provided in Appendix Equation \eqref{e:s2} with coefficient values given in \texttt{setting2\_coefficients.csv}.

To evaluate the three algorithms, we generate synthetic datasets. The following characteristics of the synthetic datasets are informed by real LS4L2 study data: (i) participant enrollment and the learning timescale, (ii) timing of decision points for each participant, and (iii) distribution of context covariates. We do not consider age and gender in the simulation.

We compare the performance of the three algorithms by computing average cumulative regret for each synthetic dataset. Informally, regret at a single decision point is the difference in the probability that the app was opened under the optimal and realized actions. Mathematically, we define cumulative regret at decision point $T$ by $R_T := \sum_{t=1}^T [.95*P(Y=1|A^*, X) + .05*P(Y=1|1-A^*, X)] - P(Y=1|\hat{A}_t, X)$ where $A^*$ is the action that maximizes $P(Y=1|X)$, $\hat{A}_t$ is the action selected by the learning algorithm at decision point $T$, and $Y$ is the binary reward. Our definition of cumulative regret reflects that the randomization probability in LS4L2 is bounded above by 0.95 to allow for continued exploration and ensure the study team can assess time-varying causal moderation with the LS4L2 data. 

The LS4L2 and simple algorithms were deployed on each of the 100 generated synthetic datasets (50 in each of the two settings).  Due to computational constraints, the complicated algorithm was deployed on 5 synthetic datasets in each setting. Average cumulative regret across the synthetic datasets was calculated for each algorithm in each setting. We additionally calculated pointwise interquartile bands for the cumulative regret for the LS4L2 and simple algorithms.  All simulations were carried out in Python, with Bayesian logistic regression models fit using the \texttt{pymc3} package \cite{pymc3} that uses Markov-Chain-Monte-Carlo (MCMC) using the no-u-turn-sampler (NUTS). 

\subsection{Simulation Results}

\begin{figure}
    \centering
    \includegraphics[width=0.9\linewidth]{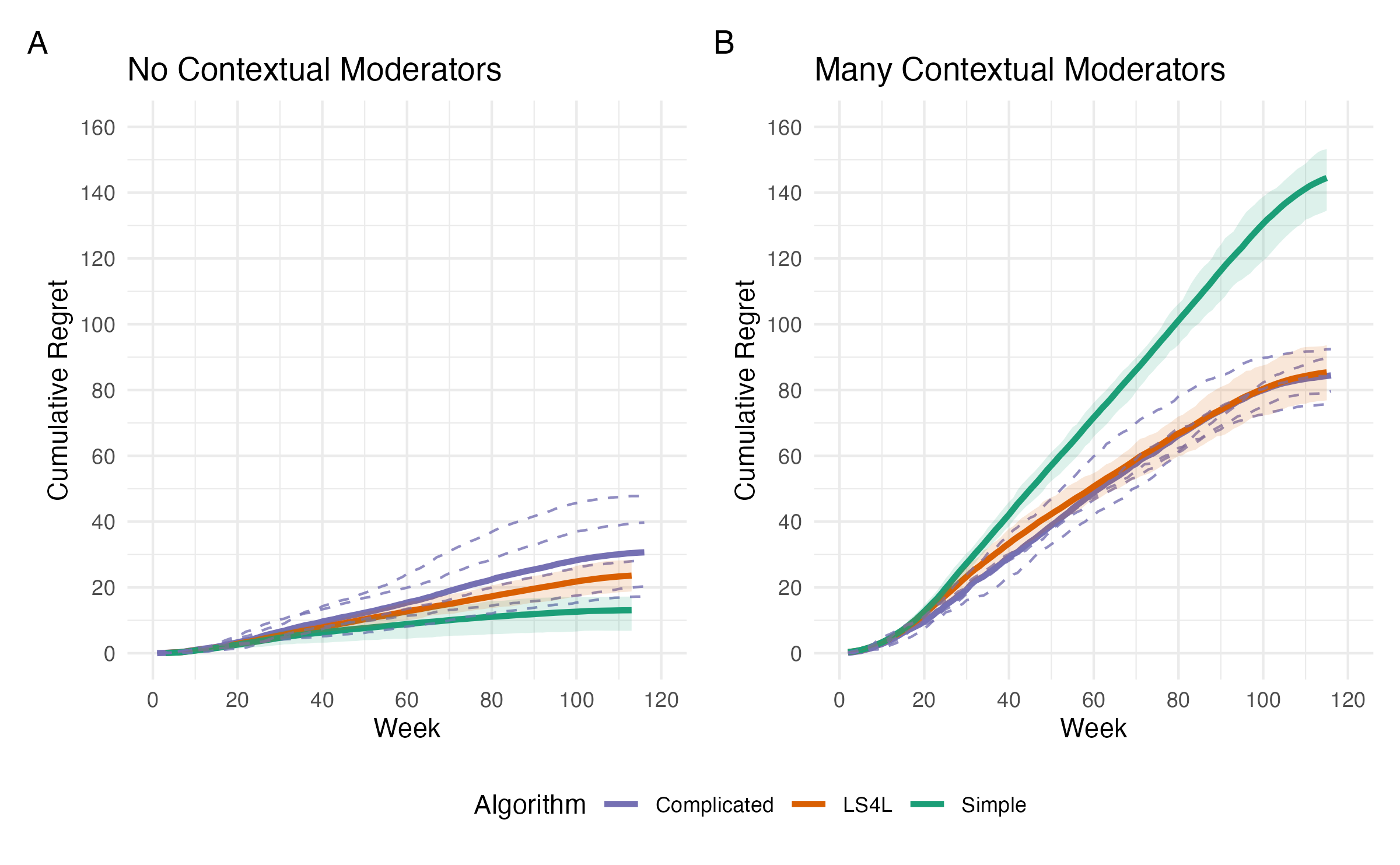}
    \caption{Average cumulative regret across simulated datasets for each algorithm and in each setting are presented, along with pointwise interquartile bands for the cumulative regret for the LS4L2 and Simple algorithms. Due to computational constraints, the Complicated algorithm was only deployed on 5 simulated datasets in each setting. The cumulative regret curves for each of these simulations are shown as dashed lines.}
    \label{f:sims}
\end{figure}

Figure~\ref{f:sims} presents the average cumulative regret for both settings.  
In setting 1 (no contextual moderators), we find the ``Simple'' algorithm led to the lowest average cumulative regret.   Our proposed LS4L2 algorithm is a compromise between the ``Simple'' and ``Complicated'' approaches (Figure \ref{f:sims}A). The superior performance of the Simple algorithm was anticipated, as it involves estimating significantly fewer parameters than the alternatives.  
In setting 2 (many contextual moderators), the LS4L2 and ``Complicated'' algorithms yield similar average cumulative regret (Figure \ref{f:sims}B), while the ``Simple'' algorithm falls behind significantly due to its inability to capture the treatment effect heterogeneity. 
The gap in average cumulative regret among the three algorithms is much smaller in setting 1 relative to the gaps we see in setting 2. 

Simulation results provide empirical support for the the study team's algorithmic design choices.  Specifically, the LS4L2 algorithm was designed to retain acceptable performance across these two settings while also accounting for computational costs. 
Even with parallel computing, the complicated algorithm took over 14 days to complete per simulation. This degree of computational complexity is cost-prohibitive in many online RL algorithms.

\section{Discussion}\label{s:discussion}

In this paper, we discuss challenges in the deployment of an online RL algorithm in clinical trials and their solutions. Our work contributes to the growing literature on frameworks for RL algorithm development in mHealth by diving into the technical tradeoffs among personalization, computation, and stability.  Specifically, we discuss how model flexibility leads to a sharp trade-off between personalization, learning, computation, and robustness. In LS4L2, our flexible model specification and use of weakly informative default priors ensured we could personalize with a computationally tractable model and reduced the risk that this model would break during the study period.  Moreover, we emphasize the distinction between the reward for the RL algorithm and behavioral task of interest.  In LS4L2, this distinction helped us think about when the reward and task were well aligned and when they were not.  This will inform future studies as we seek better rewards that align with the behavioral task of interest. 

Future work can be categorized into three directions. First, we think the issue of computation and online data storage will become even more important as study size grows.  Scalable, approximate Bayesian methods should be considered as well as federated learning techniques where parameters are shared across local models to avoid fitting the computationally expensive, hierarchical model.  Second, contextual bandit algorithms are myopic, focusing on improving the proximal outcome with no weight on delayed effects.  In mHealth, we expect most of the delayed effects to be due to habituation and/or burden.  In this case, we may consider adding a penalty (cost) to sending a message to account for these delayed effects as in~\cite{Trella_2022}.  Alternatively, we may consider using discounted reward or long-term average reward Markov Decision Process approaches~\cite{Sutton2018}. However, typically these approaches require a large amount of data to learn effective policies. Finally, often there is an intermediate outcome that is more aligned with the behavioral task.  In this case, it may be appropriate to consider the intermediate outcome and consider an episodic RL approach~\cite{pmlr-v258-gao25a}.  

\section*{Acknowledgments}

Funding source: This work was supported by the National Institutes of Health [NHLBI R61/R33 HL155498, September 2021].

\bibliographystyle{apalike}
\bibliography{bib}

\newpage

\renewcommand{\theequation}{A.\arabic{equation}}
\renewcommand{\thesection}{A.\arabic{section}}
\renewcommand{\thefigure}{A.\arabic{figure}}
\renewcommand{\thetable}{A.\arabic{table}}

\setcounter{equation}{0}
\setcounter{section}{0}
\setcounter{figure}{0}
\setcounter{table}{0}

\section*{Appendix}

\section{Rules for including age and gender in the model}
\label{sec:rules}
For baseline covariates of age and gender, we apply a slightly different set of rules to determine their inclusion.  If the data contain at least two subjects with different ages, then age is included as a main effect.  If the data contain at least four subjects with different ages, then an interaction term between age and treatment is also included in the model.  The same rules are applied to gender.  To determine if a three-way interaction term between age, gender, and treatment should be included, we check to see if at least one cell in the two-by-two table for age and gender has at least four observations (where age is temporarily converted to a binary variable by partitioning values at the mean for purposes of creating the contingency table).

\section{Simulation Design Details}
\label{app:simdetails}

\subsection{Setting 1: No Contextual Moderators}
In setting 1, the effect of treatment is constant across levels of the context covariates. The model governing the true probability of opening the app in a given context, $\boldsymbol{X} = (X_1, ..., X_p)$ and under a specific treatment, $A \in \{0, 1\}$, where $S_1 = \textbf{I}(\texttt{daysInStudy} > 56)$ and $S_2 = \textbf{I}(\texttt{daysInStudy} > 112)$:

\begin{equation}
\begin{aligned}
    logit(p_i| A_i, X_i) =& \beta_0 + \beta_1 A_i + \sum_{j = 1}^{p} \beta_{2j} X_{ij} + \sum_{j = 1}^{p} \sum_{j^{'} = j+1}^{p} \beta_{3j,j^{'}} X_{ij} X_{ij^{'}} \\& +\bigg[\beta^{(1)}_0 + \sum_{j = 1}^{p} \beta^{(1)}_{2j} X_{ij} + \sum_{j = 1}^{p} \sum_{j^{'} = j+1}^{p} \beta^{(1)}_{3j,j^{'}} X_{ij} X_{ij^{'}} \bigg] \times S_{1i} \\
    &+ \bigg[\beta^{(2)}_0 + \sum_{j = 1}^{p} \beta^{(2)}_{2j} X_{ij} + \sum_{j = 1}^{p} \sum_{j^{'} = j+1}^{p} \beta^{(2)}_{3j,j^{'}} X_{ij} X_{ij^{'}} \bigg] \times S_{2i}
\end{aligned}
\label{e:s1}
\end{equation}

 To reflect our belief that sending a notification results in participants being about 1.5 times more likely to open the smartphone app compared to those who did not receive a notification, we sampled the true treatment effect from a normal distribution with mean 0.4 and standard deviation 1/8. 

The remaining true coefficients were sampled from priors according to Table \ref{t:interactions}.
\begin{table}[h!]
\centering
\begin{tabular}{|c|c|c|c|}
\hline
\textbf{Interaction Order} & \textbf{Prior Distribution} & \textbf{Mean} & \textbf{SD} \\
\hline
1 & Normal & 0 & 1.000 \\
2 & Normal & 0 & 0.250 \\
3 & Normal & 0 & 0.0625 \\
4 & Normal & 0 & 0.0156 \\
\hline
\end{tabular}
\caption{Prior specifications by interaction order.}
\label{t:interactions}
\end{table}

The exact values of each coefficient in setting 1 can be found in \texttt{setting1\_coefficients.csv}.

\subsection{Setting 2: Many Contextual Moderators}
In setting 2, the effect of treatment is free to vary across levels of the context covariates.  The mathematical model that determines the true probability that the app is opened under a given treatment, A, in a given context, $\boldsymbol{X}$ is given below in equation \ref{e:s2}. 

\begin{equation}
\begin{aligned}
    logit(p_i|A_i, X_i) = \beta_0 + \beta_1 A_i + \sum_{j = 1}^{p} \beta_{2j} X_{ij} + \sum_{j = 1}^{p} \sum_{j^{'} = j+1}^{p} \beta_{3j,j^{'}} X_{ij} X_{ij^{'}} + \sum_{j = 1}^{p} \beta_{4j} A_i X_{ij} + \sum_{j = 1}^{p} \sum_{j^{'} = j+1}^{p} \beta_{5j,j^{'}} A_i X_{ij} X_{ij^{'}} \\
    + \bigg[\beta^{(1)}_0 + \beta^{(1)}_1 A_i + \sum_{j = 1}^{p} \beta^{(1)}_{2j} X_{ij} + \sum_{j = 1}^{p} \sum_{j^{'} = j+1}^{p} \beta^{(1)}_{3j,j^{'}} X_{ij} X_{ij^{'}}  +\sum_{j = 1}^{p} \beta^{(1)}_{4j} A_i X_{ij} + \sum_{j = 1}^{p} \sum_{j^{'} = j+1}^{p} \beta^{(1)}_{5j,j^{'}} A_i X_{ij} X_{ij^{'}} \bigg] \times S_{1i} \\
    + \bigg[\beta^{(2)}_0 + \beta^{(2)}_1 A_i + \sum_{j = 1}^{p} \beta^{(2)}_{2j} X_{ij} +  \sum_{j = 1}^{p} \sum_{j^{'} = j+1}^{p} \beta^{(2)}_{3j,j^{'}} X_{ij} X_{ij^{'}} + \sum_{j = 1}^{p} \beta^{(2)}_{3j} A_i X_{ij} + \sum_{j = 1}^{p} \sum_{j^{'} = j+1}^{p} \beta^{(2)}_{5j,j^{'}} A_i X_{ij} X_{ij^{'}} \bigg] \times S_{2i}
\end{aligned}
\label{e:s2}
\end{equation}

Setting 2 is designed so that time of the week, time of the day, and engagement level are strong moderators of treatment effect. The following specification was chosen so that the optimal treatment (i.e. the decision to send a notification) varies according to the values of contextual variables at a given decision point. During the week, daytime hours, and when engagement is low, we assume that the effect of treatment is normally distributed with mean -0.4 and standard deviation 1/8. We also assume that the difference in treatment effect when engagement is high, at night, or when it is the weekend has a normal distribution with mean .5 and standard deviation 1/16. Further, we sampled the study-period interaction with treatment from a normal distribution with mean -0.1 and standard deviation 1/32 to reflect our strong belief that treatment effect diminishes over the course of the study. As in setting 1, the remaining coefficients were sampled from priors according to Table \ref{t:interactions}

The exact values of each coefficient in setting 2 can be found in \texttt{setting1\_coefficients.csv}.

\end{document}